# Are Hidden-Variable Theories for Pilot-Wave Systems Possible?

*Louis Vervoort, 05.04.2018*

*School of Advanced Studies, University of Tyumen, Russia*

*l.vervoort@utmn.ru*

**Abstract**: Recently it was shown that certain fluid-mechanical 'pilot-wave' systems can strikingly mimic a range of quantum properties, including single particle diffraction and interference, quantization of angular momentum etc. How far does this analogy go? The ultimate test of (apparent) quantumness of such systems is a Bell-test. Here the premises of the Bell inequality are re-investigated for particles accompanied by a pilot-wave, or more generally by a resonant 'background' field. We find that two of these premises, namely outcome independence and measurement independence, may not be generally valid when such a background is present. Under this assumption the Bell inequality is possibly (but not necessarily) violated. A class of hydrodynamic Bell experiments is proposed that could test this claim. Such a Bell test on fluid systems could provide a wealth of new insights on the different loopholes for Bell's theorem. Finally, it is shown that certain properties of background-based theories can be illustrated in Ising spin-lattices.

# 1. Introduction.

Since the birth of quantum mechanics, physicists have been intrigued by its counterintuitive features, such as the collapse of the wave function, the uncertainty relations, wave-particle duality, the probabilistic nature of the theory, etc. Many attempts have been made to restore a more classic character to quantum mechanics, notably by efforts aiming at deriving the theory from a more fundamental, deeper-lying theory – maybe even a deterministic one. Such a hidden-variable theory (HVT) would contain yet unknown variables which, once integrated-out, yield quantum theory. This would after all be a familiar situation in physics: quantum mechanics would have the status of an 'effective' theory, as other theories. But this quest for sub-quantum theories, initiated by such giants as Einstein, de Broglie, Schrödinger etc., has been restrained, not only by the unprecedented precision and efficiency of quantum theory, but also by certain abstract mathematical results that seem to prove the impossibility of constructing any reasonable HVT. These are the so-called 'no-go' theorems, among which Bell's theorem [1-3] but also the Kochen-Specker theorem [4] are best known. What I call 'Bell's theorem' (as a physical, and not just mathematical, theorem) states, in short, that 'local hidden-variable theories are impossible'. It is essential here to be clear about what

'local' means: unless explicitly stated otherwise, 'local' will mean here '*only invoking not-faster-than-light interactions*' – the relevant meaning in relativity theory. Thus I will use the physics rather than the information-theoretic jargon, where 'local' has different meanings and is usually equated with 'satisfying a Bell inequality' or with 'jointly satisfying the Bell-premises' (namely Eq. (2.1a), (2.1b), (2.1c) below), or with other conditions [5]. Of course, in the context of physics the important and more general question is: *can quantum mechanics be derived from a local theory – local in the sense defined above*. But according to the received wisdom condensed in Bell's theorem, this is not possible.

But one cannot escape from following observation: hundred years after quantum mechanics' birth, and fifty years after's Bell's discovery, a large part of physicists remains skeptical about the most extraordinary claim of the standard interpretation of quantum mechanics and of Bell's theorem; namely that nature is either a-causal (quantum events arise out of nothing, have no causes) or nonlocal (superluminal interactions exist), or both. This is such a counterintuitive and from other respects unlikely prospect, that attempts to adopt a more classical picture deserve full attention – especially in view of the lasting problem of unifying quantum mechanics and relativity theory. Here we will explore such a potential revision of Bell's theorem, not of course of its mathematics, but of the physical validity of its premises ((2.1a), (2.1b), (2.1c) below). We are well aware that our arguments contain a speculative element; but the good news is that they can be tested on model systems. And we predict that in the process a wealth of new insights on the real Bell experiment will be gained – as a minimal pay-off.

Now, recent experiments go against the possibility of local HVTs, since they are generally believed to have closed the most relevant loopholes for such theories (e.g. [6-8], in these references one finds the rich history of Bell tests). These experiments use a sophisticated set-up to close e.g. the detection, locality and freedom-of-choice loopholes, notably by quickly and randomly varying analyzer settings, more generally by imposing spacelike separation between relevant events. One supplementary merit of these articles (e.g. [6-8]) is that they have sharply defined under which conditions the Bell inequality may be assumed to hold, therefore under which precise assumptions local HVTs would be eliminated. In particular, the authors of Ref. [6] specify in detail how their experiment would close the freedom-of-choice loophole: "This loophole can be closed only under specific assumptions about the origin of [the HV] λ. Under the assumption that λ is created with the particles to be measured, an experiment in which the settings are generated independently at the measurement stations and spacelike separated from the creation of the particles closes the loophole."

Here I will investigate the admissibility of local HVTs that do *not* assume that the HVs λ are 'created with the particle pair'. In particular HV models will be studied that include variables describing, not the particle pair, but a background field in which the Bell particles propagate. As is implicit in the above quote from [6], there is an argument why for such theories the freedom-of-choice loophole cannot be closed. Indeed, in a slightly more precise wording, experiments can 'close the freedom-of-choice loophole' in that they 'impose measurement independence' (MI), or the stochastic independence between the HVs λ and the left and right analyzer variables (a,b). As recalled in Section 2, all Bell inequalities assume MI, the condition (2.1c) below. Under the precise assumptions of the above quote, MI is valid (unless the universe would be superdeterministic or nonlocal). But assume now more generally that $\lambda \equiv (\lambda_0, \lambda_1, \lambda_2)$, where $\lambda_1$ are properties of a background field in the space-time neighborhood of the left analyzer setting (at angle a), $\lambda_2$ properties of the field close to the right analyzer setting (b), and $\lambda_0$ all other relevant properties. Then the conditional probability $\rho(\lambda|a,b) \equiv \rho(\lambda_0, \lambda_1, \lambda_2|a,b)$ will in general be different from the unconditional $\rho(\lambda_0, \lambda_1, \lambda_2)$, simply because $\lambda_1$ can interact with analyzer 'a' and $\lambda_2$ with analyzer 'b'. Examples of this type of correlations in known systems, hydrodynamic pilot-wave systems and spin-lattices, will be given in Section 3 and the Appendix. In Section 4 it will be argued that *in the presence of a resonant background or pilot-wave* MI-violation does not necessarily imply violation of free will or nonlocality. This suffices to reach our main conclusion, namely that in a pilot-wave system one of the premises of the Bell-inequality, namely MI, does not necessarily hold, and that therefore the Bell inequality could *possibly* be violated in such a system. This result can thus be seen as corroborating a series of articles (see in particular [9-11, 21-25, 33, 43-45, 48-49]), which have contested the physical validity of MI and therefore of the no-go theorem. As far as we know Brans [43] was the first to construct an explicit model violating MI and reproducing the quantum correlation of the Bell experiment.

In Section 4 it will be argued that in a resonant background a second premise of the Bell inequality, namely outcome independence (OI, 2.1a) can also be violated in a physically acceptable manner, i.e. compatible with locality. All depends on what the hidden-variables are supposed to represent. A background-based or pilot-wave toy model will be given that violates both MI and OI and that maximally violates the Bell inequality, while yet being non-signaling (Section 4).

To physically justify the model, it will be shown in Section 3 that the types of correlations needed (violation of MI and OI) exist in well-studied systems involving a resonant background. Indeed, this work is inspired by fluid-mechanical systems that were discovered about a decade ago by Couder, Fort and collaborators [13-15], and since then investigated in great detail also by other teams, notably by Bush and collaborators [16-18] and Gilet and collaborators [46-47]. These systems, oil droplets walking over a vibrating fluid film, exhibit a remarkable series of quantum-like features, all induced by a pilot-wave. Section 3 summarizes the essential probabilistic properties of these hydrodynamic systems; it will be shown that the long-range correlations we need massively arise here, due to resonant interaction of the droplets with a pilot-wave. In the Appendix Ising spin-lattices are investigated, which are also massively correlated through a stochastic background (namely a collection of spins in Boltzmann equilibrium). Even if spin-lattices are static and therefore not realistic as an analogue for pilot-wave HVTs, they can illustrate at least some properties of the toy model, in particular violation of MI and of the Bell inequality. It is hoped they could serve as a 1$^{st}$-order approximation of realistic HVTs. Note that spin-lattices are cellular automata, and that they thus have a simple conceptual link with 't Hooft's HVT, the Cellular Automaton Interpretation of quantum mechanics [21, 48]. 't Hooft also contests the general validity of MI.

The conclusions we arrive at can be tested, by experimental means that seem well within current reach. Indeed, any *classical* background-based system, in particular the hydrodynamic system of [13-18], that would violate a Bell inequality would corroborate our model. Since the detailed correlations in the hydrodynamic system are not known, we cannot propose a detailed experiment; but a not yet fully specified class of hydrodynamic Bell-tests can be proposed (Section 5). Finally, Section 6 discusses candidate theories for the generic model: it was recently argued (e.g. [16, 40, 42]) that the surprising quantum / hydrodynamic analogies incite one to revisit sub-quantum theories as de Broglie's pilot-wave theory [19] and its modern variants, notably stochastic electrodynamics [20], as well as other theories involving a background or pilot wave [39-41]. Preliminary ideas of the present work were presented in [23], but here a different and substantially elaborated model is given, as well as a comparison with spin-lattices.

## 2. Premises of Bell-inequalities.

Let us first recall the assumptions on which Bell-type inequalities are based. In a Bell-type experiment one measures correlations $P(\sigma_1,\sigma_2|a,b)$, i.e. joint probabilities for finding an outcome with value '$\sigma_1$' (±1) on the left particle and '$\sigma_2$' on the right particle, given that the value of the analyzer variable on the left is 'a', and 'b' on the right (the analyzer variables themselves will sometimes, when necessary for clarity, be denoted by x and y). A HV model assumes that these correlations can be explained by HVs $\lambda$ with a distribution $\rho$. In the most general stochastic setting, of which the deterministic case is but a special instance, the (minimal) premises of the Bell inequality are the following conditions, often termed 'outcome independence' (OI), 'parameter independence' (PI) (or 'no-signaling') and 'measurement independence' (MI) (e.g. [10]):

$$P(\sigma_1|\sigma_2,a,b,\lambda) = P(\sigma_1|a,b,\lambda) \qquad (OI), \qquad (2.1a)$$

$$P(\sigma_2|a,b,\lambda) = P(\sigma_2|b,\lambda) \text{ and similarly for } \sigma_1 \qquad (PI), \qquad (2.1b)$$

$$\rho(\lambda|a,b) = \rho(\lambda|a',b') = \rho(\lambda) \, . \qquad (MI) \qquad (2.1c)$$

These conditions of stochastic independence are supposed to hold for all relevant values of the variables ($\lambda$, $\sigma_1$, $\sigma_2$, x, y) appearing in the model or theory. Note that the conjunction of OI and PI is equivalent to the well-known Clauser-Horne factorability condition [28]:

$$P(\sigma_1,\sigma_2|a,b,\lambda) = P(\sigma_1|a,\lambda).P(\sigma_2|b,\lambda). \qquad (2.2)$$

Assuming the validity of (2.1a-c) one finds, using standard rules of probability calculus:

$$P(\sigma_1,\sigma_2|a,b) = \int P(\sigma_1|a,\lambda).P(\sigma_2|b,\lambda).\rho(\lambda).d\lambda. \qquad (2.3)$$

The form (2.3) leads without further assumptions to a Bell inequality:

$$X_{BI} = M(a,b) + M(a',b) + M(a,b') - M(a',b') \leq 2 \quad \forall (a,a',b,b'), \qquad (2.4)$$

where the average product

$$M(x,y) = <\sigma_1.\sigma_2>_{x,y} = \sum_{\sigma_1\sigma_2} \sigma_1.\sigma_2 P(\sigma_1,\sigma_2|x,y). \qquad (2.5)$$

OI, PI and MI are generally believed to describe *any* reasonable HVT. Jointly they are sufficient conditions for a Bell inequality to hold for the given model or theory; they are however not necessary conditions (e.g. MI may be violated in a model while the Bell inequality is still satisfied). OI and PI conjoined or the factorability condition (2.2) are often termed the 'locality condition' and are assumed to necessarily follow from Einstein locality [3]. However, in Section 4 it will be argued that OI is not necessarily a good characterization of locality in pilot-wave systems. An even more subtle premise is MI. The usual reasoning to justify this assumption goes as follows: in a local system $\lambda$ must be independent of (a,b); if not (a,b) would depend on $\lambda$ (by Bayes' rule); but that is impossible because (a,b) can be freely or randomly chosen and such free variables cannot be determined (in the probabilistic sense) by variables that determine the particle outcomes – unless one accepts a conspiratorial (superdeterministic) or a nonlocal world. Based on this classic argument [3, 9-11] one often calls MI 'freedom-of-choice'. But as mentioned above, the general validity of MI may be questioned [9-11, 21-25, 33, 43-45, 48-49]. In our view, the most cogent argument against MI is the fact that in a truly deterministic world it is not applicable [24, 49].

Another argument against MI and OI is linked to the nature of the HVs $\lambda$: do they describe the particle pair or should they be conceived more broadly ? And at what time these variables are taken? Some authors specify that the λ belong to the particles, like Bell in his original article [1] and Aspect even in his latest review [29]. But for most authors the λ are not restricted to particle properties and can or should represent *any* additional information – cf. e.g. the review article [5], and Bell in more recent articles [2, 3]. In [2], p. 56 Bell says: "It is notable that in this argument nothing is said about the locality, or even localizability, of the variable λ. […] It is assumed only that the outputs [$\sigma_1$] and [$\sigma_2$], and the particular inputs a and b, are well localized". However we will argue below that if the 'hidden variables' include a pilot-wave MI and OI are not necessarily valid.

Since the role of a background medium and of massive correlation induced by such a medium can most concretely be illustrated by the hydrodynamic experiments of Refs. [13-18], let us turn to these.

## 3. Correlations in the droplet-systems of refs. [13-18].

In a series of experiments the groups of Couder, Fort, Bush, Gilet and others have succeeded in creating oil droplets that hover over a vertically vibrating oil film. Under precise experimental conditions the droplets rapidly bounce on the oil surface, while being propelled by the surface wave they create. With such 'walking' droplets an impressive series of experiments can be performed exhibiting properties that mimic quantum phenomena, such as single particle diffraction and interference, quantization of angular momentum (the droplets can only rotate on a discrete series of radiuses), a form of tunneling, etc. [13-18, 46-47]. When the oil bath is rotated, two droplets attract each other via the surface wave they generate. An increasing rotation speed lifts the degeneracy between states, which is the analogue of Zeeman splitting [14]. Moreover these experimental analogies are backed-up by certain intriguing formal analogies[1] (e.g. [47]). For our purpose, we do not need to take into account any of the detailed mechanisms involved (for in-depth Newtonian modeling of the system, see e.g. [17]). We only need to note following general probabilistic properties (P1 – P4) of the droplet-systems:

**P1**. *The stable 'walking' regime is a probabilistic regime*. As illustrated by the complex phase diagram of possible movements [17], the walking regime occurs only in well-defined experimental conditions, i.e. for precise values of the control parameters of the system. These essentially are the frequency (f) and amplitude (A) of the external vibration, the mass and size of the droplet (m and r), the geometrical parameters of the oil film and bath ($\{d_i\}$), and the viscosities of film and droplet ($\mu_f$ and $\mu_d$). If these parameters {f, A, r, m, {$d_i$}, $\mu_f$, $\mu_d$} lie within the precise ranges of values documented by the researchers, the droplets walk horizontally; outside these ranges the movement becomes erratic and/or the droplet is captured by the film. In general, the properties measured in this regime are probabilistic: see e.g. the histogram of droplet locations in Ref. [17], Fig. 3; the multimodal probability distribution for orbital movement ([18] Fig. 10-12); etc. If certain parameters are fine-tuned deterministic trajectories can also be obtained[2]; but in general there are stochastic fluctuations, due to the non-linear (hence potentially chaotic) nature of the system and the presence of uncontrollable environmental parameters, such as air flow, temperature fluctuations etc. Note that the deterministic case is a special case of the probabilistic case, when only probabilities 0 or 1 occur.

---

[1] As one example, replacing the de Broglie wavelength in the expression of the radius of the Landau-levels by $\lambda_F$, the characteristic wavelength of the droplets' pilot-wave (the so-called Faraday wavelength), leads to radiuses that fit well to those measured on the rotating droplets [14].

**P2**. *The droplets coherently move in resonance with a periodic pilot-wave.* The droplets' movement is accompanied and induced by a structured pilot-wave, showing high degrees of periodicity and symmetry. The vibrating film gives kinetic energy to the droplet, but the bouncing droplet back-reacts, i.e. periodically hits the film and determines the shape and characteristics of a surface wave on the oil film. This pilot-wave propels the droplets horizontally over the film: while the droplets hit the wave periodically at a determined location just before the crests they receive a constant horizontal momentum. The surface field shows a high degree of symmetry; to good approximation it can be modeled by Huygens – Fresnel theory as a superposition of the circular waves created at the successive impacts.

**P3**. *In the stable regime the droplet systems are massively correlated, i.e. any system variable is potentially correlated to any other system variable, also at different spacetime points*. This is a consequence of P1 and P2. Take for instance $\lambda_1$ to be a field property (e.g. the wave height or velocity) at a certain spacetime point, and $\lambda_2$ any other field property at any other spacetime point (lying within the boundaries imposed by the experiment). Then $\lambda_1$ and $\lambda_2$ will in general be correlated, i.e.

$$P(\lambda_1|\lambda_2) \neq P(\lambda_1), \qquad (3.1)$$

due to the structure (periodicity and symmetry) of the wave. A simple and illustrative case is when $\lambda_1$ and $\lambda_2$ are heights of a circular wave at two points that are symmetrical with respect to the center (and taken at identical times); the center is here the impact point of the droplet. If the wave would be perfectly circular, and $\lambda_1$ and $\lambda_2$ have identical values, the correlation would be perfect ($P(\lambda_1|\lambda_2)$ = 1 ≠ $P(\lambda_1)$). Note that this correlation may exist even if $\lambda_1$ and $\lambda_2$ are taken at spacetime points that are spacelike separated, e.g. when they are simultaneous. Here and in the following symbols as '$\lambda_1$' and 'a' may stand for n-tuples (n-vectors). Note that P3 also holds for the correlation between wave and droplet properties. For instance, if $\lambda_0$ is a property of the droplet, say its mass, then in general

$$P(\lambda_1|\lambda_0) \neq P(\lambda_1). \tag{3.2}$$

---

[2] John W. M. Bush, private communication

This simply reflects the fact that the droplet's mass partly determines the characteristics of the surface field, as is well documented in [13-18]. $\lambda_0$ might also be a *dynamical* property of the droplet, such as its position or velocity at a given spacetime point, possibly different from the spacetime point at which $\lambda_1$ is taken. Obviously (3.2) still holds in general, due to the fact that the droplet coherently follows the structured wave. Or $\lambda_1$ and $\lambda_0$ might both be particle properties, as in [16], Fig. 4a, showing the correlation between droplet speed and location. Note, finally, that P3 also holds for the correlation between e.g. the surface field variables and the control parameters {f, A, r, m, {$d_i$}, $\mu_f$, $\mu_d$}. Also the latter parameters can be considered stochastic variables: they can be varied in certain experiments (as has been done by the experimenters). Thus we can have stochastic dependencies of the type:

$$P(\lambda_1, \lambda_2|a,b) \neq P(\lambda_1, \lambda_2) \tag{3.3}$$

where (a, b) are e.g. two variables $\in$ {$d_i$}, the dimensions of the oil bath. This reflects the fact that the properties of the pilot wave ($\lambda_1$, $\lambda_2$) strongly depend on the geometry of the bath.

**P4**. *The ubiquitous correlation of P3 also holds for a 2-droplet system.* Indeed, it has been shown that if two droplets are deposited on a vibrating film they will (always in the stable regime) create a common wave field that strongly correlates their movement. For instance, they can rotate about each other while bouncing in phase or anti-phase [14]. In this case identical z-positions of the droplets are perfectly correlated; etc. To use again a jargon from relativity theory, P4 implies in particular that in a 2-droplet system there are correlations between subsystems *that are spacelike separated*. This is not different from the 1-droplet case (cf. e.g. eq. (3.1)).

Let us now consider a case that has not been performed in the experiments, but that is a straightforward extrapolation of the latter. What would happen if two (identical) droplets could be created in the center of a (roughly symmetric) bath while receiving opposite horizontal momentum? Then one expects[3] that in the walking regime the droplets will move in opposite directions, and that underneath them a surface field with high symmetry will form which again correlates their movements (the simplest assumption is that their movements will be perfectly symmetric; but in realistic cases there will be stochastic deviations). As in eq. (3.2), the properties of the 2-droplet wave field will again depend on the droplet properties, e.g. the mass $\lambda_0$ of both identical droplets. Thus in general

$$P(\lambda_1, \lambda_2 | \lambda_0) \neq P(\lambda_1, \lambda_2) \qquad (3.4)$$

where $\lambda_1$ and $\lambda_2$ are again properties of the wave field, for instance the height ($\lambda_1$) of the surface wave at some reference point on the average trajectories of the left-moving droplets, and $\lambda_2$ the same property for the right-moving droplets. One will also have correlations of following type:

$$P(\lambda_1 | \lambda_2, \lambda_0) \neq P(\lambda_1 | \lambda_0). \qquad (3.5)$$

---

[3] John W. M. Bush, private communication

In other words $\lambda_0$ does not 'screen off' the correlations between $\lambda_1$ and $\lambda_2$; for fixed $\lambda_0$ (say mass) the field variables remain, of course, strongly correlated. This also generally holds when $\lambda_0$ stands for a set of control or 'contextual' parameters, say a $\subset \{d_i\}$, the dimensions of the oil bath:

$$P(\lambda_1 | \lambda_2, a) \neq P(\lambda_1 | a), \text{ or equivalently,} \qquad (3.6a)$$

$$P(\lambda_1, \lambda_2 | a) \neq P(\lambda_1 | a) P(\lambda_2 | a). \qquad (3.6b)$$

Fixing some contextual or control parameters does not necessarily decouple field variables, rather to the contrary.

Summarizing P1-P4, in the stable regime there potentially are correlations between any two system variables, even if these variables describe spacelike separated subsystems. P1-P4 are straightforward manifestations of the fact that the system is guided by a structured pilot wave, and that the droplets move coherently with the wave. In the next Section a generic model for a Bell experiment will be proposed in which Bell particles and analyzers interact with a background field or medium. Importantly, for that model I will only rely on probabilities of the type (3.1) – (3.6) that exist in fluid-mechanical systems.

A last remark concerns the 'path-memory' effect in the droplet-systems, studied in detail by Couder et al. [42]. The wavefield as a whole arises out of the superposition of the different waves created along the path of the droplet, so that it contains in a sense a memory of the past. Moreover the trajectory is perturbed by waves scattered by far-away obstacles through a kind of echo-location effect. Couder et al. have termed these non-local effects, but strictly speaking this is a misnomer in the present context of Bell's theorem, where 'non-local effect' means 'superluminal effect'; and of course there is nothing superluminal going on in the droplet systems. 'Delocalization' effect would be a better term. At any rate we do not rely on this path-memory effect in the model of next Section. Most importantly, when considering a hydrodynamic Bell-type experiment one can easily design it in such a way that these delocalized effects cannot come into play (Section 5).

## 4. Background-based Hidden Variable Theories. A Toy Model.

Goal is to devise a model for a Bell experiment in which the two Bell particles and analyzers interact with an (unknown) background medium or field – say the quantum vacuum, the 'zero-point field' of stochastic electrodynamics [20], an ether, etc. We do not assume any particular physical properties for this background, except the very general probabilistic features outlined above. In view of their generality it is not excluded that, besides the above fluid systems, other media or fields exhibit exactly these properties.

In a stochastic HV model for a Bell experiment one assumes that the left and right spins, $\sigma_1$ and $\sigma_2$, are stochastically determined by some $\lambda$. Here I assume that the spins are not only determined by the particle properties and the respective analyzer variables, but also by the background. The latter is supposed to interact with the particles, but also with the analyzers. Specifically, in *a background-based HV model* I assume that $\sigma_1$ [$\sigma_2$] can meaningfully be described by the probability $P(\sigma_1|\lambda_0,\lambda_1,a)$ [$P(\sigma_2|\lambda_0,\lambda_2,b)$]. Here $\lambda_1$ (an n-vector) are properties of the background field in the spacetime neighborhood of the left setting event (the analyzer angle assuming a value a); similarly $\lambda_2$ are properties of the field close to the right analyzer (b) just before or at measurement. And $\lambda_0$ are all other relevant variables that are independent on a and b, e.g. properties 'that are created with the particles to be measured' in the sense of [6] (in the simplest case these are intrinsic properties of the particles, such as mass, but in principle a broader interpretation is possible).

First assume that the factorability condition (2.2) holds, where now $\lambda \equiv (\lambda_0, \lambda_1, \lambda_2)$:

$$P(\sigma_1,\sigma_2|\lambda,a,b) \equiv P(\sigma_1,\sigma_2|\lambda_0,\lambda_1,\lambda_2,a,b) = P(\sigma_1|\lambda_0,\lambda_1,a) \; P(\sigma_2|\lambda_0,\lambda_2,b), \qquad (4.1)$$

for all values of the variables. So $\sigma_1$ only depends on a, $\lambda_0$ and $\lambda_1$, with which it is in local contact during measurement; similarly on the right. In a local background-based ('BB') model the joint probability $P(\sigma_1,\sigma_2|a,b)$ can then generally be written as follows, using (4.1):

$$P^{BB}(\sigma_1,\sigma_2|a,b) = \sum_{\lambda_0,\lambda_1,\lambda_2} P(\sigma_1,\sigma_2|\lambda_0,\lambda_1,\lambda_2,a,b) P(\lambda_0,\lambda_1,\lambda_2|a,b)$$

$$= \sum_{\lambda_0,\lambda_1,\lambda_2} P(\sigma_1|\lambda_0,\lambda_1,a) P(\sigma_2|\lambda_0,\lambda_2,b) P(\lambda_0|a,b) P(\lambda_1,\lambda_2|\lambda_0,a,b). \qquad (4.2)$$

For $\lambda_0$ we assume, by construction, that:

$$P(\lambda_0|a,b) = P(\lambda_0). \qquad (4.3)$$

Indeed, under the usual assumption of locality and absence of superdeterminism, if $\lambda_0$ is for instance an intrinsic particle property 'created at emission', and if 'a' and 'b' are randomly set at spacelike distances from the emission as e.g. in [6-8], then (4.3) must hold. Thus (4.2) becomes:

$$P^{BB}(\sigma_1, \sigma_2 | a, b) = \sum_{\lambda_0} P(\lambda_0) \sum_{\lambda_1, \lambda_2} P(\sigma_1 | \lambda_0, \lambda_1, a) P(\sigma_2 | \lambda_0, \lambda_2, b) P(\lambda_1, \lambda_2 | \lambda_0, a, b). \quad (4.4)$$

The first key property of a background-based / pilot-wave model is that MI is violated in it:

$$P(\lambda_0, \lambda_1, \lambda_2 | a, b) \neq P(\lambda_0, \lambda_1, \lambda_2 | a', b'), \quad (4.5)$$

for a ≠ a' or b ≠ b'. Since it is assumed that $\lambda_1$ directly interacts with the left analyzer at angle a, and $\lambda_2$ with the right analyzer, it immediately follows that the probability distribution for $(\lambda_1, \lambda_2)$ [and hence for $\lambda = (\lambda_0, \lambda_1, \lambda_2)$] will in general be dependent on (a,b) (cf. Appendix for examples in spin-lattices). More generally (4.5) can be understood as expressing that the field characteristics are determined by (a,b), as happens in hydrodynamic systems: cf. the correlation (3.3). Note that Eq. (4.5) can well be compatible with locality; one only needs to assume local and localized interactions for Eqs. (4.5) or (3.3) to hold.

It is instructive to first focus on a special case and assume that the analyzer settings are not varying over time, as in a static Bell experiment. Furthermore, let us assume that $\lambda_0$ takes only one fixed value. Since MI is violated, it is not a surprise that also the Bell inequality can be violated in the background model. Indeed, recently several toy models have been devised reproducing the quantum statistics via violation of MI only (other models might violate OI and/or PI): see the reviews in [5, 10, 12]. Using a result by Di Lorenzo [12] it is immediate to show that the quantum correlation, $P^{QM}$, can be recovered as a special instance of a background model that assumes only (4.1) – (4.5). The quantum correlation is:

$$P^{QM}(\sigma_1, \sigma_2 | a, b) = \frac{1}{4}[1 - \sigma_1.\sigma_2.\cos(a-b)]. \quad (4.6)$$

When $\lambda_0$ has a fixed value Eq. (4.4) reduces to a probability of the type:

$$P^{BB}(\sigma_1, \sigma_2 | a, b) = \sum_{\lambda_1, \lambda_2} P(\sigma_1 | \lambda_1, a) P(\sigma_2 | \lambda_2, b) P(\lambda_1, \lambda_2 | a, b). \quad (4.7)$$

In (4.7) all variables may be n-tuples, in particular 3-vectors. Making the substitutions $\lambda_1 \to \overline{\lambda_1}$, $\lambda_2 \to \overline{\lambda_2}$ and $(a,b) \to (\overline{a}, \overline{b})$ and assuming that $\overline{\lambda_1}, \overline{\lambda_2}, \overline{a}$ and $\overline{b}$ are unit vectors, then with the normalized choices of ref. [12]:

$$P(\sigma_1 | \overline{\lambda_1}, \overline{a}) = \frac{1}{2}(1 + \sigma_1 \overline{\lambda_1} \cdot \overline{a}),$$

$$P(\sigma_2 | \overline{\lambda_2}, \overline{b}) = \frac{1}{2}(1 + \sigma_2 \overline{\lambda_2} \cdot \overline{b}),$$

$$P(\overline{\lambda_1}, \overline{\lambda_2} | \overline{a}, \overline{b}) = \frac{1}{4} d\overline{\lambda_1} d\overline{\lambda_2} \sum_{p=\pm\overline{a}, \pm\overline{b}} \delta(\overline{\lambda_1} - \overline{p}) \delta(\overline{\lambda_2} + \overline{p}), \tag{4.8}$$

(4.7) reduces to the quantum correlation (4.6), as a simple calculation shows. Note this model is 'local' in the sense that OI and PI are obviously satisfied; yet there appears to be a hidden nonlocality in it. Indeed, the expression for $P(\overline{\lambda_1}, \overline{\lambda_2} | \overline{a}, \overline{b})$ in (4.8) can only be considered local and non-superdeterministic if the Bell experiment is static: it assumes that $\overline{\lambda_1}$ depends on $\overline{b}$ and $\overline{\lambda_2}$ on $\overline{a}$, which is only conceivable if a delocalized (but subluminal) influence is established between the left and right wings. (In the droplet systems this can occur, since the global wave field can be shaped by far-away boundary conditions.) But such a mechanism cannot work in a dynamic experiment, where such long-range interactions cannot have an effect, precluding the correlation (4.8). And indeed, it is straightforward to show that model (4.8) is 'signaling' (cf. below; not all of the conditions (4.18) are satisfied). Interestingly, several essential properties of toy model (4.8) will be seen to exist in spin-lattices.

Note that the model (4.8) not only violates MI; it does so in such a manner that

$$P(\overline{\lambda_1}, \overline{\lambda_2} | \overline{a}, \overline{b}) \neq P(\overline{\lambda_1} | \overline{a}) P(\overline{\lambda_2} | \overline{b}). \tag{4.9}$$

This is a property of non-factorability, or rather *'non-screening-off'* at the hidden-variable level: $(\overline{a}, \overline{b})$ do not screen-off the correlations between $\overline{\lambda_1}$ and $\overline{\lambda_2}$, in other words in an ensemble with fixed analyzer variables the HVs are not decoupled. If the equality sign holds in Eq. (4.9) one immediately proves that the correlation (4.7) satisfies a Bell inequality.

Beyond model (4.8), condition (4.9) will appear to be a general property of background-based / pilot-wave models. A first important hint that such models are physically allowed is that (4.9) is ubiquitous in physical systems that are compatible with free will and locality. Indeed, (4.9) is a type of correlation that exists in the droplet-systems: (4.9) is a special case of (3.6) [as can be seen by replacing in (3.6b) $\lambda_1 \to \overline{\lambda_1}, \lambda_2 \to \overline{\lambda_2}$ and $a \to (\overline{a}, \overline{b})$]. Of course, in the hydrodynamic experiments the control parameters (a,b) can be freely set, as in the Bell experiment. Therefore (3.6) and (4.9) are surely compatible with free will.

In the generic model (4.4) the non-screening-off condition is:

$$P(\lambda_1, \lambda_2 | \lambda_0, a, b) = P(\lambda_1 | \lambda_0, a, b) P(\lambda_2 | \lambda_0, \lambda_1, a, b)$$
$$= P(\lambda_1 | \lambda_0, a) P(\lambda_2 | \lambda_0, \lambda_1, b) \neq P(\lambda_1 | \lambda_0, a) P(\lambda_2 | \lambda_0, b). \tag{4.10}$$

The first equality follows from the product rule of probability calculus; the second from a locality assumption that $\lambda_1$ only depends on a (not on b), and similarly in the right wing; and the last inequality expresses the fact that $\lambda_1$ and $\lambda_2$ are conditionally correlated. If the equality sign holds in (4.10), (4.4) is of the Bell-Clauser-Horne type and satisfies a Bell inequality. A correlation of the type (4.10) exists in droplet-systems, but also in spin-lattices (cf. Appendix).

Let us now turn to the validity of OI in pilot-wave systems; also this assumption is not so obvious as is usually believed. First, note that the decorrelations (2.1a) and (2.2) are assumptions which are surely allowed by probability theory, but for which the literature ultimately provides only an intuitive justification (cf. e.g. [1-3, 5, 9-11]). In the end (2.1a) and (2.2) are just probabilistic assumptions, and within physics no proof exists that OI could not be violated in supercorrelated systems as the droplet-systems. Inspecting the definition (2.1a) of OI, it becomes clear that all depends on what one denotes by the HVs '$\lambda$':

$$P(\sigma_1|\sigma_2,a,b,\lambda) = P(\sigma_1|a,b,\lambda). \quad \text{(OI)} \quad (2.1a)$$

Consider for instance a Bell experiment on the droplet systems (involving two correlated droplets, cf. Section 5) and take $\lambda = \lambda_0$, some intrinsic droplet property as mass, and $\sigma_1$ and $\sigma_2$ some dynamical droplet properties taken at measurement (say the z-component of their position above a reference plane), then it is quite obvious that in general OI will *not* be valid. Indeed (a,b, $\lambda_0$) are contextual parameters influencing the pilot-wave properties ($\lambda_1$ and $\lambda_2$); which are in general correlated (Section 3). Since the droplets ($\sigma_1$ and $\sigma_2$) resonantly follow the pilot-wave (property P2, Section 3), also the latter properties will be correlated in general. E.g. $P(\sigma_1|a,b, \lambda_0)$ may be = 0.5 while $P(\sigma_1|\sigma_2,a,b, \lambda_0)$ may be ≈ 1 if $\sigma_1 \approx \sigma_2$, due to the resonant movement in a symmetric surface field. $P(\sigma_1|\sigma_2,a,b, \lambda_0) = 1$ may hold in the ideal case of a perfectly symmetric surface field. It would thus appear that OI is questionable, and that the core of the locality condition (2.2) in background-based systems is not OI but PI, since PI must necessarily be satisfied.

Accepting violation of OI, we have in general, instead of (4.1):

$$P(\sigma_1,\sigma_2|\lambda,a,b) \equiv P(\sigma_1,\sigma_2|\lambda_0,\lambda_1,\lambda_2,a,b)$$
$$= P(\sigma_1|\lambda_0,\lambda_1,a)\, P(\sigma_2|\sigma_1,\lambda_0,\lambda_2,b) \neq P(\sigma_1|\lambda_0,\lambda_1,a)\, P(\sigma_2|\lambda_0,\lambda_2,b). \quad (4.12)$$

A next question is then: can one construct a background-based toy model (i.e. satisfying (4.4), (4.10) and (4.12)) that violates the Bell inequality while being non-signaling ? This is indeed possible, as will now be shown. If the arguments above are correct, such a model would be compatible with locality (not involve superluminal influences) and free will. We have now:

$$P^{BB}(\sigma_1,\sigma_2|a,b) = \sum_{\lambda_0} P(\lambda_0) \sum_{\lambda_1,\lambda_2} P(\sigma_1|\lambda_0,\lambda_1,a) P(\sigma_2|\sigma_1,\lambda_0,\lambda_2,b) P(\lambda_1|\lambda_0,a) P(\lambda_2|\lambda_0,\lambda_1,b), \quad (4.13)$$

where the probabilities should satisfy the following normalization conditions:

$$\sum_{\lambda_0} P(\lambda_0) = 1, \qquad (4.14a)$$

$$\sum_{\sigma_1} P(\sigma_1 | \lambda_0, \lambda_1, a) = 1, \qquad \forall (\lambda_0, \lambda_1, a), \qquad (4.14b)$$

$$\sum_{\sigma_2} P(\sigma_2 | \sigma_1, \lambda_0, \lambda_2, b) = 1, \qquad \forall (\sigma_1, \lambda_0, \lambda_2, b), \qquad (4.14c)$$

$$\sum_{\lambda_1} P(\lambda_1 | \lambda_0, a) = 1, \qquad \forall (\lambda_0, a), \qquad (4.14d)$$

$$\sum_{\lambda_2} P(\lambda_2 | \lambda_0, \lambda_1, b) = 1, \qquad \forall (\lambda_0, \lambda_1, b). \qquad (4.14e)$$

For the proof it is sufficient to make judicious choices for the probabilities in (4.13-14); in particular let us assume that $\lambda_0$ has one fixed value and that $\lambda_1$ and $\lambda_2$ only assume two values, which can be set to 1 and 2:

$$\lambda_1, \lambda_2 = 1, 2. \qquad (4.15)$$

Eq. (4.13) can then be written without loss of generality:

$$P^{BB}(\sigma_1, \sigma_2 | a, b) = \sum_{\lambda_1, \lambda_2} P(\sigma_1 | \lambda_1, a) P(\sigma_2 | \sigma_1, \lambda_2, b) P(\lambda_1 | a) P(\lambda_2 | \lambda_1, b). \qquad (4.16)$$

Make now following normalized choices, satisfying (4.14):

$$P(\lambda_1 = 1 | a) = 0.5 = P(\lambda_1 = 1 | a')$$
$$P(\sigma_1 = +1 | \lambda_1 = 1, a) = 1 = P(\sigma_1 = +1 | \lambda_1 = 2, a'); \quad P(\sigma_1 = +1 | \lambda_1 = 1, a') = 0 = P(\sigma_1 = +1 | \lambda_1 = 2, a)$$
$$P(\sigma_2 = +1 | \sigma_1 = +1, \lambda_2 = 1, b) = 1 = P(\sigma_2 = +1 | \sigma_1 = +1, \lambda_2 = 2, b)$$
$$P(\sigma_2 = +1 | \sigma_1 = +1, \lambda_2 = 1, b') = 0; \quad P(\sigma_2 = +1 | \sigma_1 = +1, \lambda_2 = 2, b') = 1$$
$$P(\sigma_2 = +1 | \sigma_1 = -1, \lambda_2 = 1, b) = 0; \quad P(\sigma_2 = +1 | \sigma_1 = -1, \lambda_2 = 2, b) = 1$$
$$P(\sigma_2 = +1 | \sigma_1 = -1, \lambda_2 = 1, b') = 0; \quad P(\sigma_2 = +1 | \sigma_1 = -1, \lambda_2 = 2, b') = 1$$
$$P(\lambda_2 = 1 | \lambda_1 = 1, b) = 1 = P(\lambda_2 = 1 | \lambda_1 = 2, b); \quad P(\lambda_2 = 1 | \lambda_1 = 1, b') = 0; P(\lambda_2 = 1 | \lambda_1 = 2, b') = 1.$$

$$(4.17)$$

These choices (model BB-1) are summarized in Table 1:

| x | y | $\lambda_1$ | $\sigma_1$ | $\lambda_2$ | $\sigma_2$ |
|---|---|---|---|---|---|
| a | b | 1 | +1 | 1 | +1 |
|   |   | 2 | -1 | 1 | -1 |
| a' | b | 1 | -1 | 1 | -1 |
|   |   | 2 | +1 | 1 | +1 |
| a | b' | 1 | +1 | 2 | +1 |
|   |   | 2 | -1 | 1 | -1 |
| a' | b' | 1 | -1 | 2 | +1 |
|   |   | 2 | +1 | 1 | -1 |

**Table 1**. Background-based model BB-1 assumes that the analyzer variables (x and y) are equiprobable and independent (P(x,y) = P(x)P(y) = 0.25 $\forall$ (x,y)) and that P($\lambda_1$=1|x) = 0.5, $\forall$ x; the other variables have values with a conditional probability 0 or 1 (cf. (4.17)).

For the analyzer variables it is assumed P(x) = P(y) = 0.5 and P(x,y) = 0.25, $\forall$ (x,y), as is usual in Bell-experiments. This model is semi-deterministic since besides the analyzer variables (x,y) and $\lambda_1$ all other variables have only values with a conditional probability 0 or 1. Thus if (x,y) = (a,b), Table 1 shows that $\sigma_1$ and $\sigma_2$ have equal sign, therefore M(a,b) = <$\sigma_1\sigma_2$>$_{a,b}$ = +1. Similarly M(a',b) = M(a,b') = +1 and M(a',b') = –1, implying $X_{BI}$ = 4: model BB-1 maximally violates the Bell inequality. □

Note, and this is essential, that the model does not allow superluminal signaling[4]. Indeed, Table 1 shows there is no dependence between the relevant left- and right-wing variables, namely between y and $\sigma_1$; y and $\lambda_1$; y and ($\sigma_1$, $\lambda_1$); and similarly for x. If e.g. y and $\sigma_1$ would be dependent, then measuring (many times) $\sigma_1$ would inform Alice about Bob's setting choice, allowing superluminal signaling. In detail, as can be seen in Table 1, model BB-1 satisfies following non-signaling conditions (these are not all independent, but that is not important here):

$$P(\sigma_1|x,b) = P(\sigma_1|x,b') \quad \forall (\sigma_1, x)$$
$$P(\lambda_1|x,b) = P(\lambda_1|x,b') \quad \forall (\lambda_1, x)$$
$$P(\sigma_1,\lambda_1|x,b) = P(\sigma_1,\lambda_1|x,b') \quad \forall (\sigma_1, \lambda_1, x)$$
$$P(\sigma_2|a,y) = P(\sigma_2|a',y) \quad \forall (\sigma_2, y)$$
$$P(\lambda_2|a,y) = P(\lambda_2|a',y) \quad \forall (\lambda_2, y)$$
$$P(\sigma_2,\lambda_2|a,y) = P(\sigma_2,\lambda_2|a',y) \quad \forall (\sigma_2, \lambda_2, y). \qquad (4.18)$$

Now, a basic assumption of the model is (4.10), non-screened-off MI-violation, which is for fixed $\lambda_0$:

$$P(\lambda_1,\lambda_2|a,b) = P(\lambda_1|a) P(\lambda_2|\lambda_1,b) \neq P(\lambda_1|a) P(\lambda_2|b). \qquad (4.19)$$

The reader unaware of (4.18) could be tempted to conclude that (4.19) necessarily is nonlocal (allows superluminal signaling), based on following erroneous reasoning: (4.19) shows that $\lambda_2$ depends on b and that $\lambda_1$ depends on $\lambda_2$; hence $\lambda_1$ must depend on b, which is nonlocal. But probabilistic dependence is not necessarily transitive; and Table 1 (satisfying in particular the second of the conditions (4.18)) indeed proves that $\lambda_1$ does *not* depend on b. The reason why dependence is not transitive here is that $\lambda_1$ and $\lambda_2$ do not cause each other; they are caused by the action of their respective analyzer (a[b] partly determining $\lambda_1$[$\lambda_2$]) *and* by a common cause, namely the action of the particle pair simultaneously creating (or partly influencing) $\lambda_1$ and $\lambda_2$. In sum there is no direct causal path from $\lambda_1$ to b.

---
[4] A related model presented in [23] is local in the sense of Clauser-Horne, but is signaling.

More generally, going beyond model BB-1, it seems possible to physically justify our basic assumptions (4.10) and (4.19) (MI-violation) and (4.12) (OI-violation) by reference to the droplet-systems, also in a dynamic Bell-experiment. As already stated, these types of correlations are encountered in the hydrodynamic systems: e.g. (4.10) is a correlation of the type (3.5) and (3.6) (replace e.g. a in (3.6) by (a, b, $\lambda_0$)). In the droplet systems (4.10) can be interpreted in a simple manner: in an ensemble with fixed droplet masses ($\lambda_0$) and fixed control parameters (a,b), there exists in general a correlation between field properties ($\lambda_1$, $\lambda_2$) – as manifestly follows from the experiments [13-18] (property P4, Section 3). We thus can explain (4.10) and (4.19) in a real Bell experiment by the hypothesis that each particle pair leaving the source generates, or is accompanied by, a structured background wave showing high symmetry and periodicity; hence $\lambda_1$ and $\lambda_2$, properties of the background wave taken at measurement at (symmetric) points close to the detectors, can be strongly correlated. Likewise (4.12) can also be justified by the fact that the particles, as the droplets, resonantly move in a structured background field. Finally, it seems that there is no reason why varying analyzer settings, as in the advanced experiments [6-8], would have to destroy the correlations we need. To maintain (4.10) and (4.12), one simply needs to assume that the switching does not totally disrupt the structure in the pilot-wave, in other words the long-range correlations. A simple physical picture in which this happens is when the analyzers only slightly influence the properties $\lambda_1$, $\lambda_2$ of the pilot-wave (on their respective sides) compared to the particles themselves ($\lambda_0$); more generally (in model BB-1 the dependence between $\lambda_2$ and b is strong) the analyzers should not totally randomize the correlation between $\lambda_1$ and $\lambda_2$. This seems a harmless assumption not violating any known physical law. In the droplet systems the droplet properties ($\lambda_0$) strongly determine the pilot-wave; a small object in the bath, characterized by a (varying) parameter 'a', can well be assumed to have a smaller influence – small enough for sufficient correlation between $\lambda_1$ and $\lambda_2$ to persist[5].

Thus a background or pilot-wave model based on (4.10) and (4.12) can potentially violate a Bell inequality; we have argued it is compatible with locality (subluminal interactions) and free will. Clearly, this is a 'no no-go' claim based on a toy model, not a proof that there exist real background systems violating a Bell inequality. In the spirit of recent work by Popescu and others [36-37], model BB-1 is "even more nonlocal than quantum mechanics, yet fully consistent with relativity". Following these authors, an interesting question is whether there are physical principles that would *limit* the value of $X_{BI}$ in more realistic background models. This question may be more easily answered here, since background models have a physical interpretation.

---

[5] Note that some influence of a on $\lambda_1$ and on $\sigma_1$ and/or of b on $\lambda_2$ and on $\sigma_2$ is still required, else the Bell inequality is always satisfied. A corollary of above analysis is that the background system envisaged here might need to be nonlinear, in the sense that tiny influences of one parameter (a) on another ($\lambda_1$) cause a macroscopically detectable difference in outcome in a third quantity ($\sigma_1$). Interestingly, this is again observed in the droplet systems, which exhibit such exponential sensitivity on initial conditions, as shown in [15, 17]. The importance of non-linearity in the framework of local HVTs was already noted in [25].

At this point a rather obvious question arises, namely whether the above model involving "particles + a pilot wave", can be generalized to just "particles as (stochastic) classical fields". Indeed, in the background model I considered the interaction of Bell particles with a stochastic background, where in the simplest interpretation $\lambda_0$ is a property of the particle and $\lambda_{1(2)}$ of the field. But can one not suppose that *all* properties $\lambda_0, \lambda_1, \lambda_2$ describe a stochastic classical field, and *that the particles are singularities of the field*, $\lambda_0$ describing these singularities ? (One will note the analogy of this picture with quantum field theories, in which particles are excited states of the delocalized vacuum state.) It seems that also in this picture we can have the essential correlations (4.10) and (4.12). Then one would come to the almost discomfortingly simple conclusion that as soon as the HVs λ are associated with extended fields, Bell's theorem is in jeopardy – under the conditions discussed for the background model. But is this not a perfect prelude to quantum field theories ?

In the Appendix it is shown that some of the probabilistic properties considered above actually exist in Ising spin-lattices (for a fuller treatment see [38]). Just as droplet-systems, spin-lattices are highly correlated through a stochastic background (namely the spins in between the 'test' and 'control' spins in Fig. 2). They are not a dynamic system as a Bell experiment and are not realistic templates for pilot-wave HVTs, but they nevertheless allow to nicely illustrate several properties of the generic background model, in particular non-screened-off MI-violation. They are 'local' in the sense of satisfying OI and PI (the Clauser-Horne factorability), but they are signaling, and thus nonlocal in this sense. Maybe their description can be generalized to a dynamical situation; it is well known that related lattice-gas models exist for the Navier-Stokes equation [35].

## 5. Hydrodynamic Bell experiments.

In view of the lasting mysteries surrounding quantum mechanics and Bell's theorem, it seems that the droplet-systems, as macroscopic pilot-wave systems, offer a unique opportunity for advancing the debate on several fronts. Above it was argued that in such systems MI and OI are questionable and may not hold. Note that MI, OI and PI are sufficient but not necessary conditions for the Bell inequality; so even if for instance MI would not be valid in some pilot-wave system there is no guarantee that a Bell inequality is violated. All depends on the system in question (including all boundary and measurement conditions) and on the precise strength of the correlations (4.13) it generates. Although we do not know the numerical values of these correlations for the droplet systems [13-18], the walkers are obviously excellent candidates to test the generic model. Let us therefore present some guidelines for such experiments. These seem particularly relevant in view of the difficulty to devise genuinely loophole-free experiments on quantum systems [6-8, 50]: this should be much easier in the macroscopic realm. In order to close e.g. the locality loophole one just needs to consider the sonic speed in the medium, not the light speed.

Clearly, Bell's reasoning leading to the Bell inequality does nowhere rely on the fact that $\sigma_1$ and $\sigma_2$ in (2.1) are spins; *any* bi-partite, classical, local system on which free-willed experimenters perform a Bell experiment should satisfy a Bell inequality according to [1-3]. Thus $\sigma_1$ and $\sigma_2$ may – a priori – represent *any* property of the droplets. This greatly opens the parameter space of the experiments that can be envisaged. A few general remarks on such a hydrodynamic Bell experiment can be made without specifying[6] ($\sigma_1$, $\sigma_2$, a, b); a conceptual scheme of such a test is given in Fig. 1.

The first challenge is to generate pairs of correlated droplets that move roughly in opposite directions (but this should be possible: see P4 in Section 3). Then, in order to implement the model of Section 4, some measurement devices characterized by parameters x (left) and y (right) should interact with the fluid bath, thus determining the pilot-wave properties in their neighborhood and by the same token some dynamical droplet property ($\sigma$). The interaction should not be so strong as to totally randomize and destroy the structure in the pilot-wave for all values of the variables. Besides judiciously choosing which properties ($\sigma_1$, $\sigma_2$, x, y) to measure and to control, all control parameters {f, A, r, m, {$d_i$}, $\mu_f$, $\mu_d$,…} will have to be optimized in order to generate fine-tuned correlations. (This may not be more difficult than what was achieved in the pioneering experiments [13-18]: their authors have doubtlessly performed a great number of experiments before finding the conditions and observables showing quantum-like behavior, in other words quantum-like correlations.) The first interesting type of experiments could be static: then one has to determine on an ensemble of droplet-pairs the average products M(x,y) in (2.5) in four consecutive experiments corresponding to the four combinations (x,y). Of course, in such a static experiment one has, besides those discussed above, an additional resource for violating a Bell inequality, namely the delocalized effects that exist in the droplet-systems (cf. last paragraph of Section 3).

---

[6] There is some incentive to choose for $\sigma$ the hydrodynamic equivalent of spin, which might suggest to use a rotating bath, or rotating or magnetic droplets. Indeed, the existence of hydrodynamic 'spin states' has already been suggested [16]. But it is wise to start with the much simpler experiment suggested below.

In the ideal, dynamic test one activates (x,y) just before ($\sigma_1,\sigma_2$) are measured, in such a manner that x[y] cannot influence $\sigma_2$[$\sigma_1$] in time (taking the droplet and sound speeds in the background medium into account); this to close the locality loophole as in experiments [6-8]. This can be achieved by measuring for each droplet pair the (simultaneous) time of emission from a source ($T_0$), by activating (x,y) at time $T_0 + T_1$, and by (non-intrusively) measuring ($\sigma_1,\sigma_2$) at $T_0+T_1+\varepsilon$ where $\varepsilon$ is small enough to prevent that the perturbation induced by (x,y) can travel to the other side.

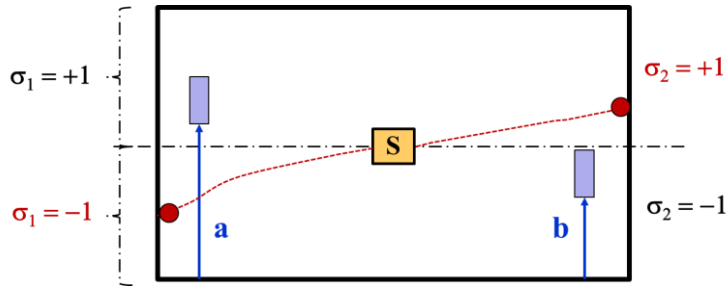

Fig. 1. Conceptual scheme of a Bell-type experiment in droplet-systems (S = source).

In Fig. 1 I took for simplicity $\sigma = +1$ if the droplet hits the side-wall above the symmetry plane of the bath; as an example a and b can be positions of some small objects, e.g. pins pushed in the bath, ideally during the 'flight' of the droplets just before measurement. A more realistic case may be to take ($\sigma_1,\sigma_2$) as the instantaneous deflections of the droplets at $T_0+T_1+\varepsilon$ (+1[−1] if the angle with respect to a reference line is positive [negative], again much as in Fig. 1). But again, Fig. 1 should merely be seen as a conceptual scheme.

Model BB-1 can be a source of inspiration: it suggests to choose (b, b') in such a manner that when y = b, $\lambda_2$ is not dependent on $\lambda_1$, while when y = b', $\lambda_2$ is (perfectly) correlated with $\lambda_1$. This suggests to take as measurement system one that can disrupt the correlation between $\lambda_1$ and $\lambda_2$ when y has a specific value (b). Model BB-1 also suggests performing the experiment in the stochastic regime, in which the droplets follow stochastic trajectories that are not straight lines. Then violation of OI is a possible resource for violating a Bell inequality (it is well-known that in deterministic systems OI is always satisfied [10]).

Even independently of our model, comparing the static and dynamic case will allow to investigate a wealth of interesting questions related to the quantum version of the Bell experiment. In general, the role of loopholes can be studied much more easily in the fluid system. For instance, if the Bell inequality would be violated in the static experiment but not in the dynamic one, interesting information on the precise dynamics of some delocalized effect could be obtained. But also the relevance and meaning of the still quite mysterious detection, fair sampling and coincidence loopholes [50] can be studied much more easily, simply because the particles can be non-intrusively observed and 'killed in flight'. We emphasize this is a phenomenal advantage over the photon experiments, where one largely moves in the dark; hence surprises are not excluded. Finally, note that all these loopholes correspond to what could be called, with caution, 'generalized non-local effects'. In nonlinear, chaotic systems as the droplet systems such effects, *even if arbitrarily weak*, could be enhanced and lead to violation of the Bell inequality, as was shown in [25].

## 6. Existing attempts at theories.

A natural question is whether the background-based theories envisaged here are already represented by existing (be it not yet finalized) theories, aiming at completing quantum mechanics via hidden background variables. The first candidates that come to mind are Madelung's hydrodynamic interpretation of quantum mechanics [31], and, especially, Louis de Broglie's pilot-wave theory [19]. Further recent and noteworthy attempts, inspired by the droplet systems, are [39-41]. Maybe the most elaborated theory is stochastic electrodynamics [20], a variant of de Broglie's pilot-wave theory. Even if these theories are not final, it has recently been argued that in view of new experimental data [13-18] they deserve renewed attention [42, 16]. In de Broglie's theory a quantum particle resonantly interacts with its pilot-wave at the Compton frequency $\omega_C = mc^2/\hbar$ (the Zitterbewegung); at the same time it is guided by a monochromatic pilot-wave in real space characterized by the de Broglie wavelength $\lambda_B = h/p$ [16]. Qualitatively this composed dynamics quite remarkably corresponds to the movement of the droplets (where the role of the de Broglie wavelength is taken by the Faraday wavelength); it is also reflected in the generic background model. In Madelung's theory as well as in variants of Bohm's theory [32] particles are dragged by a 'Madelung fluid' while undergoing Brownian deviations around the average streamlines; the latter provide the hidden background variables. Stochastic electrodynamics, investigated since the 1960ies by several researchers starting from Marshall, Boyer, de la Pena, Cetto and others (cf. review [20]), allows to recover several quantum features and equations based on one main ingredient: a stochastic 'zero-point field' (ZPF), resonantly interacting with particles and exchanging energy with them. Thus these theories involve a background field in the sense of the model of Section 4. Moreover it seems they might allow for the long-range correlations this model invokes. Indeed, in stochastic electrodynamics entanglement and *apparent* non-locality arise through '*common resonance modes*' of the particles ([20] p. 248), quite similarly as in the model of Section 4.

## 7. Conclusion

More than fifty years after its discovery, the interpretation of Bell's theorem remains highly debated, for the simple reason it leaves us the choice between two utterly unpleasing worldviews, involving either a-causality (things happen without any reason, out of the blue) or superluminal interactions. From one point of view, it is almost curious that one interpretation – termed 'superdeterminism', even if 'determinism' (the view that all events have causes) suffices as a term [24] –, has so few adepts, since it allows to solve the problem. According to this deterministic model Bell's theorem should be interpreted differently – namely as proving that MI (measurement independence) is not valid at the quantum level. Technically, determinism taken seriously amounts to non-validity of MI. We have argued here that hydrodynamic pilot-wave systems [13-18] offer a unique opportunity to test whether MI and the Bell inequality are valid at the macroscopic scale. In these systems there may be an additional ground why MI is violated, in a manner that is clearly compatible with free will (Section 4).

More precisely, we provided a background toy model violating MI and OI (outcome independence) and maximally violating the Bell inequality, while yet being non-signaling. We argued that all types of correlations needed in this model arise in the hydrodynamic pilot-wave systems of refs. [13-18]. This suggests that this toy model is compatible with locality (subluminal interactions) and free will.

As stated in Ref. [6], MI can only be imposed, and therefore the freedom-of-choice loophole closed, under the assumption 'that λ is created with the particles to be measured'. In the toy model of Section 4 the λ do not only describe (particle) properties at emission, but also a resonant background in the spacetime neighborhood of the detection events. Of course, one could say that such pilot-wave HVTs rely on another type of nonlocality, or rather holism, since they invoke delocalized fields. But the difference is clear: this is physics as usual, and not spooky action-at-a-distance. In the intuitively simplest picture, directly inspired by the droplet-systems, entanglement and *apparent* nonlocality can arise in such theories when particles coherently move in resonance with a periodic wave (or maybe, by extension, when they are singularities of such a field).

Finally, we emphasized that hydrodynamic Bell experiments could provide decisive new insights in Bell's theorem, also independently of our model. Indeed, there is a continuing debate on the relevance of loopholes invalidating Bell's theorem. It is much easier to experimentally close or open these loopholes in the hydrodynamic systems, so that their interpretation can be studied in detail.

**Acknowledgements**. I would like to thank John Bush, Marc Fleury, Lorenzo Maccone, Ward Struyve, and especially Scott Glancy for highly instructive discussions. The article is dedicated to a mentor since a decade, Mario Bunge.

## Appendix. A Bell-type experiment on spin-lattices

Let us suppose Alice and Bob share an ensemble of identical spin-lattices containing 10 spins (all $\sigma_i = \pm 1$), all at the same temperature T (Fig. 2). Also suppose Alice and Bob perform a Bell-type experiment on this ensemble by measuring $\sigma_1$, $\sigma_a$ (Alice's wing) and $\sigma_2$, $\sigma_b$ (Bob's wing).

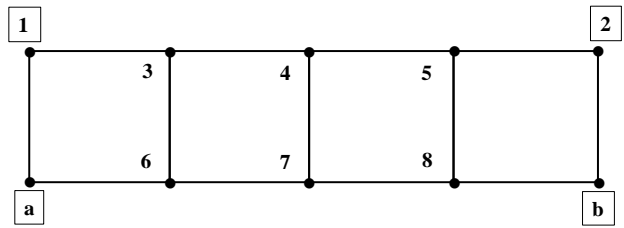

Fig. 2. 10 spins on a lattice

The Hamiltonian is the classical spin-1/2 Ising Hamiltonian:

$$H(\theta) = -\sum_{i,j} J_{ij}\sigma_i\sigma_j - \sum_i h_i\sigma_i \ . \qquad (5.1)$$

Here $\theta$ is a 10-spin configuration ($\sigma_a,\sigma_b,\sigma_1,\ldots \sigma_8$), the $h_i$ are local magnetic fields, and the $J_{ij}$ are the interaction constants between $\sigma_i$ and $\sigma_j$, as usual assumed to be zero beyond nearest neighbors. Thus the (Coulomb) interaction between the spins is local and localized; at the same time it is 'transmitting' in that the elements of the lattices 'feel' each other over long distances. Concomitantly spin-lattices are massively correlated: every spin is correlated with every other spin in the lattice [30]. This implies in particular that the system is signaling: not every condition in (4.18) is satisfied. Finally it is assumed that the probability of a given spin configuration is the usual Boltzmann probability:

$$P(\theta) = e^{-\beta H(\theta)}/Z, \text{ with } Z = \sum_\theta e^{-\beta H(\theta)}, \text{ the partition function.} \qquad (5.2)$$

It is then straightforward to calculate the Bell probabilities $P(\sigma_1,\sigma_2/\sigma_a,\sigma_b)$ and the inequality (2.5) where the role of a (b) is now taken by $\sigma_a$ ($\sigma_b$) (take a $\equiv$ b $\equiv$ +1 and a' $\equiv$ b' $\equiv$ −1). The role of the 'background' HVs is assumed by the intermediate spins: $\lambda \equiv \sigma_\lambda \equiv (\sigma_3,\sigma_4,\ldots\sigma_8)$.

As an example, if one assumes that all $J_{ij}$ are equal (all $J_{ij} = J$) and all $h_i = 0$ then one obtains (the sums $\sum_{i,j}$ run over the 13 first-neighbour pairs):

$$Z P(\sigma_1,\sigma_2,\sigma_a,\sigma_b) = \sum_{\sigma_3\sigma_4\ldots\sigma_8} e^{-\beta H(\theta)} = \sum_{\sigma_3\sigma_4\ldots\sigma_8} e^{\beta J \sum_{i,j}\sigma_i\sigma_j} = \sum_{\sigma_3\sigma_4\ldots\sigma_8} \prod_{i,j} e^{\beta J \sigma_i \sigma_j}$$

$$= \sum_{\sigma_3\sigma_4\ldots\sigma_8} \prod_{i,j} [\cosh(\beta J \sigma_i\sigma_j) + \sinh(\beta J \sigma_i\sigma_j)]$$

$$= \sum_{\sigma_3\sigma_4\ldots\sigma_8} \prod_{i,j} [\cosh(\beta J) + \sigma_i\sigma_j \sinh(\beta J)] = \sum_{\sigma_3\sigma_4\ldots\sigma_8} (\cosh(\beta J))^{13} \prod_{i,j}[1 + \sigma_i\sigma_j \tanh(\beta J)]$$

$$= \sum_{\sigma_3\sigma_4\ldots\sigma_8} \alpha \prod_{i,j}[1 + K\sigma_i\sigma_j]$$

$$= \alpha(1 + K\sigma_1\sigma_a)(1 + K\sigma_2\sigma_b) \sum_{\sigma_3\sigma_4\ldots\sigma_8}(1 + K\sigma_1\sigma_3)(1 + K\sigma_a\sigma_6)\ldots(1 + K\sigma_5\sigma_2)(1 + K\sigma_8\sigma_b)$$

$$= \alpha(1 + K\sigma_1\sigma_a)(1 + K\sigma_2\sigma_b) \times$$

$$\sum_{\sigma_3\sigma_4\ldots\sigma_8}[1 + K(\sigma_1\sigma_3 + \sigma_a\sigma_6 + \ldots) + K^2(\sigma_1\sigma_3\sigma_a\sigma_6 + \sigma_1\sigma_3^2\sigma_6 + \ldots) + \ldots + K^{11}\sigma_1\sigma_a\sigma_3^3\sigma_4^3\sigma_5^3\sigma_6^3\sigma_7^3\sigma_8^3\sigma_2\sigma_b]. \quad (5.3)$$

Here $\alpha \equiv (\cosh(\beta J))^{13}$ and $K \equiv \tanh(\beta J)$. The only non-zero terms are those in which all $\sigma_i$ appearing as indices (namely $\sigma_3, \sigma_4,\ldots, \sigma_8$) are squared. The lowest-order terms in which this occurs are $K^3\sigma_1\sigma_3^2\sigma_6^2\sigma_a$ and $K^3\sigma_2\sigma_5^2\sigma_8^2\sigma_b$. These terms correspond to a path linking the nodes 1-3-6-a and 2-5-8-b respectively (cf. Fig. 2); the power of K corresponds to the number of segments in the path. To retain all non-zero terms, we thus have to count 1) all direct (i.e. not self-intersecting) paths linking nodes 1 and 2, 1 and a (and 2 and b), 1 and b (and 2 and a) and a and b; 2) all closed loops (such as 3-4-7-6-3); and all products of such paths that have no segments in common (such as 1-3-6-a and 4-5-8-7-4). This leads to:

$$Z P(\sigma_1,\sigma_2,\sigma_a,\sigma_b) = \alpha(1 + K\sigma_1\sigma_a)(1 + K\sigma_2\sigma_b)2^6 \times$$

$$\times \{ 1 + (K^3 + K^5 + 2K^7)(\sigma_1\sigma_a + \sigma_2\sigma_b) + (K^4 + 3K^6)(\sigma_1\sigma_2 + \sigma_a\sigma_b) +$$

$$+ (K^6 + 3K^8)\sigma_1\sigma_2\sigma_a\sigma_b + (3K^5 + K^7)(\sigma_1\sigma_b + \sigma_2\sigma_a) + 2K^4 + K^6 \}. \quad (5.4)$$

Using the same procedure we find:

$$Z P(\sigma_a,\sigma_b) = \sum_{\sigma_1\sigma_2\ldots\sigma_8} e^{\beta J \sum_{i,j}\sigma_i\sigma_j} = \alpha \sum_{\sigma_1\sigma_2\ldots\sigma_8} \prod_{i,j}[1 + K\sigma_i\sigma_j]$$

$$= \alpha 2^8[1+\sigma_a\sigma_b(K^4+10K^6+5K^8)+4K^4+3K^6+5K^8+3K^{10}]. \tag{5.5}$$

Dividing (5.3) by (5.5) we obtain $P(\sigma_1,\sigma_2 | \sigma_a,\sigma_b)$. For instance,

$$P(+,+|+,+) = \frac{(1+K)^2[2K^3+4K^4+8K^5+8K^6+6K^7+3K^8]}{2^2[1+5K^4+13K^6+10K^8+3K^{10}]}. \tag{5.6}$$

This implies that for a lattice with homogeneous interactions $J_{ij} = J = 1$ and $\beta = 1$, $P(+,+|+,+) = 0.95$. $X_{BI}$ can then numerically be evaluated via (2.4-5). E.g., for $J=1=\beta$, $X_{BI} = -0.667$, a value that satisfies the Bell inequality. In the weak-interaction limit $K \ll 1$ one finds $X_{BI} \approx -2.K^2$, satisfying the Bell inequality. In the strong-interaction limit $K \gg 1$ one obtains $X_{BI} \approx 1$, again satisfying (2.4).

Importantly, while above outcomes are obtained for all $J_{ij} = J$ and all $h_i = 0$, numerical simulation shows that the result is more interesting if one varies the $J_{ij}$ and $h_i$ ($\neq 0$) over the lattice. In that case *the BI can be strongly violated*, for a broad interval of values for $\beta$, $h_i$, $J_{ij}$. For instance, for $\beta = 1$, $h_i \in \{-1, 1, 3\}$, $J_{ij} \in \{1, 2, 3, 4\}$, and keeping left-right symmetry in the lattice, one finds that $X_{BI} = 2.87$ at its local maximum [38]. This exceeds $2\sqrt{2} \approx 2.83$, the Tsirelson bound and the value for the singlet state in the original Bell experiment. In the smallest hexagonal lattice one finds, by an amusing coincidence, $X_{BI}^{max} = 2.82843 = 2\sqrt{2} + 3.10^{-6}$.

Even if some lattices violate the Bell inequality, it is possible to prove that they satisfy OI and PI [38] (as also follows from the theory of Markov fields [34]). *At the same time all lattices violate MI*; thus MI-violation is the resource for violation of the Bell inequality. In [38] it is further shown that non-screening-off properties as (3.5) and (4.10) hold in the lattice of Fig. 2, e.g. if one takes $\lambda_1 \equiv (\sigma_3,\sigma_6)$, $\lambda_2 \equiv (\sigma_5,\sigma_8)$ and $\lambda_0 \equiv \sigma_4$. This actually follows directly from the theory of Markov random fields.

Clearly, to make this study fully relevant one would have to go beyond a static Bell experiment; it is an open question whether Ising-like (lattice-gas) models can be adapted to the dynamical phenomena considered in Section 4, notably to describe a pilot-wave[7]. Still, classic spin-lattices already exhibit several intriguing similarities with the generic model of Section 4 and can illustrate some of the abstract correlations considered there, notably 'non-screened-off MI-violation' leading to violation of the Bell inequality.

---

[7] The Ising Hamiltonian could also be assumed for moving particles, as in lattices gases, which can be described by Ising-like Hamiltonians and which can simulate the Navier-Stokes equation. However the Boltzmann probability (5.2) is only valid in equilibrium. Note also that the Ising Hamiltonian can be seen as a 1st-order approximation (it consists of the first two terms of a Taylor expansion in the spins).